\documentclass[twcolumns]{aa}
\usepackage{graphicx}
\usepackage{hyperref}
\usepackage{xcolor}
\usepackage[varg]{txfonts}

\newcommand{\degs}{\ensuremath{{\rm deg^2}}}
\newcommand{\MICE}{\texttt{MICEcat}}
\newcommand{\frate}{\ensuremath{f_{\rm rate}}}
\newcommand{\dthr}{\ensuremath{\hat{d}^{\rm thr}_{L}}}
\newcommand{\hu}{\ensuremath{\rm{km\, s^{-1} Mpc^{-1}}}}

\usepackage[normalem]{ulem}

\bibpunct{(}{)}{;}{a}{}{,}
\nolinenumbers
\begin{document}
   \title{Investigating the impact of galaxies' compact binary hosting probability for gravitational-wave cosmology}

   % \subtitle{I. Overviewing the $\kappa$-mechanism}

   \author{G. Perna \thanks{Just to show the usage
          of the elements in the author field}
          \inst{1,2}
          \and
          S. Mastrogiovanni\inst{3}\and A. Ricciardone\inst{4,5,1}
          }

\institute{Dipartimento di Fisica e Astronomia ``Galileo Galilei'', Universit\`a degli Studi di Padova,\\ Via Marzolo 8, I-35131, Padova, Italy \and
INFN, Sezione di Padova,\\ Via Marzolo 8, I-35131, Padova, Italy 
\and 
INFN, Sezione di Roma, I-00185, Roma, Italy 
\and
Dipartimento di Fisica ``Enrico Fermi'', Universit\`a di Pisa, \\ Largo Bruno Pontecorvo 3, Pisa I-56127, Italy
\and
INFN, Sezione di Pisa, \\Largo Bruno Pontecorvo 3, Pisa I-56127, Italy
}

   % \institute{Institute for Astronomy (IfA), University of Vienna,
   %            T\"urkenschanzstrasse 17, A-1180 Vienna\\
   %            \email{wuchterl@amok.ast.univie.ac.at}
   %       \and
   %           University of Alexandria, Department of Geography, ...\\
   %           \email{c.ptolemy@hipparch.uheaven.space}
   %           \thanks{The university of heaven temporarily does not
   %                   accept e-mails}
   %           }

   \date{Received ---; accepted ----}

\abstract
{With the advent of future-generation interferometers a huge number of gravitational wave (GW) signals is expected to be measured without an electromagnetic counterpart. Although these signals do not allow a simultaneous measurement of the redshift and the luminosity distance, it is still possible to infer cosmological parameters. In this paper, we focus on the systematic biases that could arise from mismodeling the GW host probability when inferring the Hubble constant ($H_0$) with GW dark sirens jointly with galaxy catalogues. We discuss the case in which the GW host probability is a function of galaxies' luminosity and redshift as it has been predicted by state-of-the-art compact binary coalescences (CBCs) synthetic catalogues. We show that, in the limiting case in which the analysis is done with a complete galaxy catalog covering a footprint of $\sim 10$ \degs, mismatching the host probability in terms of galaxy's luminosity will introduce a bias on $H_0$. In this case, the magnitude of the bias will depend on the distribution of the Large-Scale Structure over the line-of-sight. Instead, in the limit of a complete wide-field of view galaxy catalog and GW events localized at $\mathcal{O}({\rm Gpc})$ distance, mismatching the redshift dependence of the GW hosting probability is more likely to introduce a systematic bias.
}


\maketitle

\section{Introduction}

The first observation of a gravitational wave (GW) signal in 2015 \cite{LIGOScientific:2016aoc} opened a new complementary way to access information about our Universe, both on the astrophysical and the cosmological sides. On the astrophysical side, after the detection of a further 90 GW events, it has become feasible to infer not only information about the individual sources emitting GWs but also about the mass and spin distributions of the entire population of compact binaries \cite{KAGRA:2021duu,LIGOScientific:2020kqk}. 
On the cosmological side, GWs can be used to probe the cosmic expansion \cite{1986Natur.323..310S,Holz:2005df,PhysRevD.74.063006} independently to other probes such as Supernovae \citep{Galbany:2022zir} and the Cosmic Microwave Background (CMB) \citep{Planck:2018vyg}. This is possible thanks to the fact that GWs directly provide a measurement of the source luminosity distance. Therefore, if provided with a source redshift estimation, GW sources can be used to probe the cosmic expansion. Unfortunately, GWs alone cannot provide a direct measurement of the redshift since it is degenerate with the source masses. A direct estimation of the redshift can be obtained if an electromagnetic counterpart (EM) to the GW source is observed. The EM counterpart could allow the identification of the host galaxy for which it is typically possible to measure the redshift with spectroscopic observations. Indeed, this was the case of the binary neutron star merger GW170817 \cite{LIGOScientific:2017vwq,LIGOScientific:2017ync} for which it has been possible to estimate $H_0=70^{+12}_{-8}$ km/s/Mpc at 68.3 \% credible interval (C.I.). Since then, no additional GW event has been observed with an EM counterpart.

As first argued by \citet{1986Natur.323..310S}, even in the absence of an electromagnetic counterpart, GW sources can still be employed to measure the cosmic expansion (dark sirens method). There are several techniques proposed in the literature to use dark sirens for cosmology such as the source-frame mass method \citep{Chernoff:1993th, Taylor:2011fs, Farr:2019rap, Ezquiaga:2020tns, Mastrogiovanni:2021wsd, Mukherjee:2021rtw, Leyde:2022orh, Ezquiaga:2022zkx, Karathanasis:2022rtr} and cross-correlations with large-scale structure tracers \citep{Oguri:2016dgk, Mukherjee:2019wcg, Mukherjee:2020hyn, Bera:2020jhx}. In this work, we focus on the ``galaxy catalogue'' technique first proposed by  \citet{1986Natur.323..310S}. This technique consists of assigning a statistical redshift measurement by using the GW source localization and all the galaxies reported in a given survey \citep{DelPozzo:2011vcw, Chen:2017rfc, Gray:2019ksv, Leandro:2021qlc, Gray:2021sew, 2023AJ....166...22G}.
This is possible since every astrophysical GW event is expected to be hosted in a galaxy\footnote{Unless it is sourced by primordial black holes formed at early Universe ages \citep{2018CQGra..35f3001S} or it results from the merger of binaries with large kick velocities \citep{2016PhRvL.117a1101G}. Both these types of events are expected to be very rare.}. 
Although the galaxy catalogue method is less precise inferring $H_0$ than a single event with EM counterpart, it is expected to play a relevant role for GW cosmology given the number of dark sirens detections as demonstrated by its several current applications \cite{LIGOScientific:2018gmd,DES:2019ccw,LIGOScientific:2019zcs,DES:2020nay,Palmese:2021mjm,LIGOScientific:2021aug,Finke:2021aom,mastrogiovanni2023novel,2023arXiv231205302B}. Furthermore, with the advent of future generation interferometers such as LISA \cite{LISACosmologyWorkingGroup:2022jok}, Cosmic Explorer (CE) \cite{Evans:2021gyd} and Einstein Telescope (ET) \cite{Maggiore:2019uih,Branchesi:2023mws}, a much higher number of GW signals are expected to be detected, further emphasizing the importance of such analysis. 

A central topic of the galaxy catalogue method is how to describe for each galaxy the probability of hosting a GW event. The simplest choice might be to assume that every galaxy has an equal probability to host a GW event, even though this is not a physically motivated assumption as we observe that the overall Star Formation Rate (SFR), and hence the compact objects produced, correlates with several galactic properties \citep{2014ARA&A..52..415M}. Furthermore, as shown in \cite{Hanselman}, assuming equal host probabilities can lead to biases in the estimated value of $H_0$. As such, current analyses include a hosting probability based on the intrinsic luminosity of the galaxies, namely the brighter the galaxy, the more likely it is to host a GW event (some recent predictions for host galaxies properties are discussed in \cite{Vijaykumar:2023bgs, 2020MNRAS.495.1841A}). It has been shown in \citep{Gray:2019ksv} that luminosity weights can boost the $H_0$ precision, however, there is no current study about the \textit{systematics} on $H_0$ that these assumptions can introduce.

In this work, we study in detail how modelling the GW host probability can introduce systematics when inferring $H_0$. We build our study on the simulation approach presented in \citet{2023AJ....166...22G}. Starting from a simulated Universe (\MICE) \citep{2015MNRAS.453.1513C}, we populate GW events in each galaxy following a hosting probability model obtained from synthetic catalogues of CBCs \citep{2020MNRAS.495.1841A}. We then use different selection criteria for the GW events and different GW hosting probability models to study the possible presence of biases for the $H_0$ inference.
This work is organized as follows. We start in Section \ref{Sec::Bayesian_Framework} describing the statistical framework we consider in our analysis. In Section \ref{Sec::Data_Properties} we present the simulated Universe considered for the simulation we argue about the GW data generation technique. Finally in Section \ref{Sec::Results} we discuss the results of our analysis, while in Section \ref{Sec::Conclusions} we report our conclusions.

\section{Statistical framework}
\label{Sec::Bayesian_Framework}

In this section we describe the statistical framework employed for our study. The posterior on $H_0$ inferred from the detection of $N_{\rm GW}$ events from a collection of data $\{x\}$, can be obtained using the Bayes theorem (e.g. \cite{sivia2006data})
\begin{equation}
    p(H_0|\{x\}) \propto \mathcal{L}(\{x\}|H_0) p(H_0)\,,
\end{equation}
where $p(H_0)$ is a prior and $\mathcal{L}(\{x\}|H_0)$ the likelihood. The likelihood describes the GW detection as an inhomogeneous Poisson process in the presence of selection biases \citep{Mandel:2018mve,Vitale:2020aaz}. Neglecting the information on the binary masses, for the galaxy catalogue technique the likelihood can be written as \cite{2023AJ....166...22G}
\begin{eqnarray}
    \mathcal{L}(\{x\}|H_0) &\propto& e^{-N_{\rm exp}(H_0)}[N_{\rm exp}(H_0)]^{N_{\rm GW}} \nonumber \times \\ && \prod_i^{N_{\rm GW}} \frac{\int dz d\Omega \mathcal{L}_{\rm GW}(x_i|d_{\rm L}(z,H_0),\Omega)p_{\rm CBC}(z,\Omega)}{\int dz d\Omega  P_{\rm det}^{\rm GW}(z, H_0,\Omega)p_{\rm CBC}(z,\Omega)}\, ,
    \label{eq:fund}
\end{eqnarray}
where $N_{\rm exp}$ is the expected number of GW detections, $z$ the source redshift, $\Omega$ the sky location,  $\mathcal{L}_{\rm GW}(x_i|d_{\rm L}(z,H_0), \Omega)$ the GW likelihood (i.e. how well we can measure the luminosity distance and sky position), $P_{\rm det}^{\rm GW}(z, H_0, \Omega)$ the GW detection probability and $p_{\rm CBC}(z,\Omega)$ the probability of finding CBC at redshift $z$ and sky position $\Omega$.  By assuming that the GW likelihood is reasonably constant in a given sky area $\Omega_{\rm loc}$, and that the detection probability does not vary in this sky area, i.e. $P_{\rm det}^{\rm GW}(z, H_0,\Omega) \approx P_{\rm det}^{\rm GW}(z, H_0)$, Eq.~\ref{eq:fund} can be rewritten dropping the sky-localization dependency and defining $p_{\rm CBC, loc}(z) \propto p_{\rm CBC}(z,\Omega_{\rm loc})$. In the remainder of the paper we will refer to $p_{\rm CBC, loc}(z)$ with simply $p_{\rm CBC}(z)$ for brevity reasons. Later in this section, we will explain how to construct $p_{\rm CBC}(z)$ taking into account a non-trivial GW hosting probability. Eq.~\ref{eq:fund} can be marginalized analytically on $N_{\rm exp}$ \cite{Fishbach:2018edt} by assuming a flat-in-log prior thus obtaining the identity
\begin{equation}
    \mathcal{L}(\{x\}|H_0) \propto  \prod_i^{N_{\rm GW}} \frac{\int dz \mathcal{L}_{\rm GW}(x_i|d_{\rm L}(z,H_0))p_{\rm CBC}(z)}{\int dz  P_{\rm det}^{\rm GW}(z, H_0)p_{\rm CBC}(z)}\,.
    \label{eq:fund_scale}
\end{equation}

For this work, we model the GW likelihood using the same toy model of \citet{2023AJ....166...22G},
\begin{equation}
\label{Eq::SOLikelihood}
    \mathcal{L}_{\rm GW}(\hat{d}_L^i|d_L(z,H_0)) = \frac{1}{\sqrt{2 \pi}A d_L (z,H_0)} e^{- \frac{(\hat{d}_L^i-d_L(z,H_0))^2}{2 A^2 d^2_L(z,H_0)}}\,
\end{equation}
where we identify the data chunk $x_i$ as an ``observed'' luminosity distance $\hat{d}_L^i$ on which we apply a selection cut. In Eq. above, $A=0.2$ is chosen to mimic a typical error budget on the luminosity distance of $20\%$. As argued in \cite{2023AJ....166...22G}, it is important to apply the selection cut strictly on the observed luminosity distance to avoid introducing a systematic bias due to the incongruent statistical model, see \citep{Essick:2023upv} for a general detailed discussion about physical and unphysical statistical models. The detection probability is defined for each ``true'' $d_L$ by integrating over all the observable data sets $\hat{d}_L$. Following \citep{2023AJ....166...22G}, we assume that GWs are detected if their observed luminosity distance is lower than a certain horizon $\hat{d}_L^{\rm thr}$, namely 
\begin{eqnarray}
    P_{\rm det}^{\rm GW} (d_L) &=& \int_{-\infty}^{\hat{d}_L^{\rm thr}} \mathcal{L}_{\rm GW} (\hat{d}_L|d_L(z,H_0)) d\hat{d}_L \nonumber \\
    &=& \frac{1}{2} \left[1 + {\rm{erf}}\left(\frac{d_L(z,H_0)-\hat{d}_L^{\rm thr}}{\sqrt{2}A d_L(z,H_0)}\right)\right].
\end{eqnarray}
This simple prescription of the GW detection horizon is nearly equivalent to considering a selection criteria based on the signal-to-noise ratio (SNR). In fact, for GW sources, the SNR is proportional to the inverse of the luminosity distance. Before proceeding to the construction of a galaxy-informed $p_{\rm CBC}(z)$, let us clarify that this type of statistical framework does not apply to real data. In a realistic scenario, the GW detection probability will depend on other parameters such as the detector masses of the GW signals. Moreover, we are not able to define a model that approximates $x_i$ with ``observed'' quantities. As such, real data analysis techniques such as \citet{2023arXiv230517973M}, use parameter estimation samples generated from data $x_i$ in synergy with an injection campaign to correct selection biases. For all these reasons, the computational complexity  does not allow for an extensive and straightforward Monte Carlo study on how the GW hosting probability translates to a $H_0$ bias.

\subsection{Building the compact binary coalescence probability}

We model $p_{\rm CBC}(z)$ following \citet{mastrogiovanni2023novel,2023JCAP...12..023G,2023arXiv231205302B}
\begin{equation}
    p_{\rm CBC}(z) = \frac{\int dM f_{\rm rate}(z,M) p_{\rm cat}(z,M)}{\int dM dz f_{\rm rate}(z,M) p_{\rm cat}(z,M)},
    \label{eq:pcbc}
\end{equation}

where $f_{\rm rate}$ is a function that parametrizes the probability that a galaxy with redshift $z$ and absolute magnitude $M$ is likely to host a GW signal, while $p_{\rm cat}(z,M)$ is the distribution of galaxies in terms of redshift and absolute magnitude in a given sky pixel\footnote{Let us remind the reader that this is a consequence of the fact that we have assumed the GW likelihood uniform in a given sky area.}. This parametrization of $p_{\rm CBC}(z)$ is an extension to the one provided in \citet{2023AJ....166...22G} that considered only a dependence of the CBC rate from the galaxy redshift. Moreover, this parameterization requires that the redshift and apparent magnitude (from which the absolute magnitude is computed) be accurately measured, otherwise a prior term on the distribution of absolute magnitudes in the Universe should be included. 
In principle, we could extend Eq.~\ref{eq:pcbc} to include any galactic property. Here we just focus on absolute magnitude (i.e., intrinsic luminosity) as it is the usual galactic property used to model the GW hosting probability. 
Before describing the modelization of $f_{\rm rate}(z,M)$, let us briefly comment on Eq.~\ref{eq:pcbc}. If all the galaxies are equally likely to host a GW source, then $f_{\rm rate}(z,M)$ is constant and 
\begin{equation}
    p_{\rm CBC}(z) = \int dM p_{\rm cat}(z,M)=p_{\rm cat}(z).
\end{equation}
In other words, the distribution of possible CBC hosts will conincide to the distribution of galaxies over the line-of-sight. However, if $f_{\rm rate}(z,M)$ is not uniform, this will not be the case.

We model the rate function using a model similar to \citet{mastrogiovanni2023novel}, namely
\begin{equation}
    f_{\rm rate}(z,M) =\psi_{\rm CBC}(z)\, 10^{-0.4 \,\epsilon\, (M-M_*)},
    \label{eq:fMR}
\end{equation}
where $M_*$ is the Schechter function knee value for the galaxies' luminosity distribution.  $\psi_{\rm CBC}(z)$ is a function that we compute numerically to parameterize the CBC merger rate as a function of redshift (see Sec.~\ref{Sec::Data_Properties}) and
\begin{equation}
    10^{-0.4\, \epsilon\, (M-M_*)}=\left(\frac{L}{L_*}\right)^\epsilon, 
\end{equation}
a term that introduces a power-law dependence for the host probability, which depends on the galaxies' intrinsic luminosity. This simple parametrization is inspired from \citet{2020MNRAS.495.1841A}, where it is argued, starting from synthetic simulations of CBCs, that the CBC merger rate has a dependence on the galaxies' redshift (due to the different star formation rate and time delays) and galaxies' luminosity (due to the total stellar mass). 

Finally, let us discuss how we practically calculate and employ $p_{\rm cat}(z,M)$. Assuming that we are provided with a \textit{complete} galaxy catalog reporting $N_{\rm gal}$ galaxies with a collection redshifts $\{\hat{z}_{\rm g}\}$ and apparent magnitudes $\{\hat{m}_{\rm g}\}$, then following \citet{2023AJ....166...22G,mastrogiovanni2023novel} we have
\begin{equation}
    p_{\rm cat} (z,M) \approx \frac{1}{N_{\rm gal}} \sum_i^{N_{\rm gal}} p(z|\hat{z}^i_g) \, p(m(M,H_0,z)|\hat{m}_{g,*}^i)\,,
\end{equation}
where the terms indicated with $p(\cdot)$ are the single posteriors on each galaxy \textit{true} redshift and apparent magnitude. In order to better focus on the systematics introduced by the modelling of $f_{\rm rate}(z,M)$, we assume that we have perfectly measured redshift and apparent magnitude of the galaxies. These are reasonable assumptions for galaxies whose redshit is estimated with spectroscopic observations. When the errors on the galaxies' redshift and apparent magnitude are small,
\begin{equation}
    p_{\rm cat} (z,M) \approx \frac{1}{N_{\rm gal}} \sum_i^{N_{\rm gal}} \delta(z-\hat{z}^i_g) \,\delta(m(M,H_0,z)-\hat{m}_{g,*}^i)\,.
    \label{eq:pcat}
\end{equation}
In practice, we never compute $p_{\rm cat} (z,M)$. In fact by substituting Eq.~\ref{eq:pcat} in Eq.~\ref{eq:pcbc} and later considering the hierarchical likelihood in Eq.~\ref{eq:fund_scale}, with the GW likelihood model in Eq.~\ref{Eq::SOLikelihood}, it is possible to show that
\begin{equation}
        \mathcal{L}(\{x\}|H_0) \propto \prod_i^{N_{\rm GW}} \frac{\sum_i^{N_{\rm gal}} \mathcal{L}_{\rm GW}(\hat{d}^i_L|d_{\rm L}(\hat{z}_g^i,H_0))f_{\rm rate}(\hat{z}_g^i,M(\hat{m}^i_g,H_0,\hat{z}_g^i))}{\sum_i^{N_{\rm gal}} P_{\rm det}^{\rm GW}(\hat{z}_g^i, H_0)f_{\rm rate}(\hat{z}_g^i,M(\hat{m}^i_g,H_0,\hat{z}_g^i))}\,.
        \label{eq:final}
\end{equation}

\section{Simulation setup and catalogs description}

\label{Sec::Data_Properties}
We provide an overview of the data products and simulation procedures used for this study. In Sec.~\ref{SubSec::MICECat} we describe the \texttt{MICECat} catalogue, a simulated Universe used to assign the GW galaxy hosts. In Sec.~\ref{SubSec::GW_Sim} we discuss the procedure to generate the GW signals starting from the galaxies list and in Sec.~\ref{SubSec::Sub_Cataloghs} we discuss some of the features of the GW catalogues generated from the simulation.

\subsection{Description of MICEcat and GW host probability model}
\label{SubSec::MICECat}

In this work, we use \texttt{MICECat}, a galaxy catalogue generated from the MICE Grand Challenge lightcone simulation \citep{2015MNRAS.448.2987F,2015MNRAS.453.1513C,2015MNRAS.447.1319F,2015MNRAS.447..646C,2015MNRAS.447.1724H}, i.e. a full N-body simulation of the Universe containing about 70 billion dark matter particles. \texttt{MICEcat}  has been generated using the following set of cosmological parameters, $\Omega_{\rm m} = 0.25$, $\sigma_8 = 0.8$, $n_{\rm s} = 0.95$, $\Omega_{\rm b} = 0.044$, $\Omega_\Lambda = 0.75$, $h = 0.7$ for a fiducial $\Lambda$CDM model. The lightcone simulation has been performed in the redshift range $0 < z < 1.4$, and using a hybrid halo occupation distribution and halo abundance matching prescriptions to populate friends-of friends dark matter halos, whose masses are greater than $2.2 \cdot 10^{11} M_\odot/h$ \citep{2015MNRAS.453.1513C}. \texttt{MICEcat} reports the true redshift and the true absolute magnitude $M_r$ in the infrared band (r-band) for all the galaxies. The catalogue covers only 1/8 of the full sky in the redshift range and is complete up to redshift 1.4.

We model the GW host probability for each event using the parametrization for $f_{\rm rate}$ in Eq.~\ref{eq:fMR}. The redshift dependence of the CBC rate function is obtained with the procedure used in \citet{Bellomo:2021mer}. We start from the phenomenological expression provided by Madau-Dickinson for star formation rate (SFR) in \citet{2014ARA&A..52..415M},
\begin{equation}
    \mathcal{R}_{\rm SFR}(z) \propto \frac{(1+z)^{\lambda_1}}{1+\left( \frac{1+z}{\lambda_2} \right) ^{\lambda_1+\lambda_2}}\,,
\end{equation}
with $\lambda_1 = 2.7$, $\lambda_2 = 2.9$, valid for the redshift range $0\leq x \leq 8$. We assume that there is no cosmologically significant time delay between the formation of the stellar progenitor and the compact object, this is motivated since compact objects are born from massive stars that are expected to live tens of Myrs. We then, use a simple model to account for the time delay in compact object formation and merger, the merger rate can be written as
\begin{align}
    {\psi}_{\rm CBC}(z) \propto \int_{t_{\rm d,min}}^{t(z)}  \mathcal{R}_{\rm SFR}(z(t_{\rm{d}}))p(t_{\rm{d}}) dt_{{\rm d}}\,,
\end{align}
with $t_{\rm d,min} = 50$ Myr\footnote{See \cite{Bellomo:2021mer} for additional details on the integration boundaries.} and $p(t_{\rm d}) \propto t_{\rm d}^{-1}$. We verify that, after this step, the redshift dependence of the merger rate goes like $(1+z)^{1.82}$ consistently with current observations \citep{KAGRA:2021kbb}. We find that a good fit for the merger rate, after accounting for the time delay, yields
\begin{equation}
\label{Eq::Merger_Rate_Delay}
\psi_{\rm CBC}(z) \propto\frac{1}{1+z} \frac{(1+z)^{1.82}}{1+(\frac{z}{2})^{3.82}}\,,
\end{equation}
where the $1/(1+z)$ factor is introduced to account for the time dilation between the source frame and the detector frame due to the expansion of the Universe.
The CBC rate dependence from the galaxies' intrinsic luminosities has not yet been measured. Therefore, we use a model calibrated on the astrophysical simulations of CBC formation performed in \citet{2020MNRAS.495.1841A}.  In this study, the authors provide a fit to the GW hosting probability as a function of the galaxy luminosity in the K-band (which is expected to be a reasonable tracer of the galaxy's total stellar mass). 
\begin{figure}[t!]
    \centering
    \includegraphics[width=0.5\textwidth]{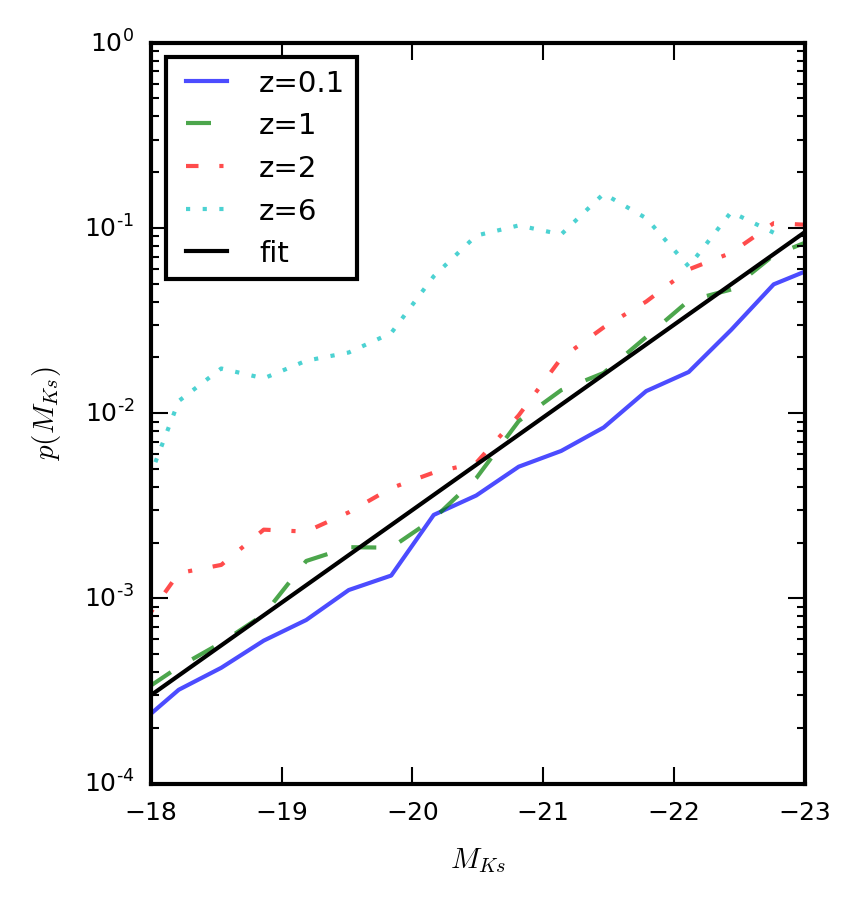}
    \caption{The plot reports the GW host probability as a function of the galaxies's absolute magnitude in the K-band. Different lines indicate the host probability at different redshifts. The GW host probability has been extracted from \citet{2020MNRAS.495.1841A}.}
    \label{Fig::MCARtale_Fit}
\end{figure}
In Fig.~\ref{Fig::MCARtale_Fit}, we report the GW host probability as a function of absolute magnitude in the K-band, $M_{K}$ provided by \citet{2020MNRAS.495.1841A}. As it can be seen from the plot, the GW host probability is not expected to strongly evolve for redshifts $\leq 2$, a limit that is above the GW events that we consider detectable in our simulation (see later). Therefore, for our case study, we assume that the GW host probability in terms of galaxy luminosity does not evolve with redshift. The best-fitting $\epsilon$ (c.f. Eq.~\ref{eq:pcbc}) in Fig.~\ref{Fig::MCARtale_Fit} is 2.25, that corresponds to GW hosting probability in terms of luminosity
\begin{equation}
    p(L) \propto L^{\frac{9}{4}},
\end{equation}
and we show in Fig.~\ref{Fig::MCARtale_Fit} in comparison with the numerical values provided by \citet{2020MNRAS.495.1841A}.
Hence, under the approximations done so far, the final merger rate model we consider results
\begin{equation}
    f_{rate}(z,L) \propto L^{\frac{9}{4}}\frac{1}{1+z}\,\frac{(1+z)^{1.82}}{1+\left(\frac{z}{2}\right)^{3.82}}\,.
\end{equation}

\subsection{Simulation of the GW catalogue}

\label{SubSec::GW_Sim}
\begin{figure}
    \centering
    \includegraphics[scale=0.4]{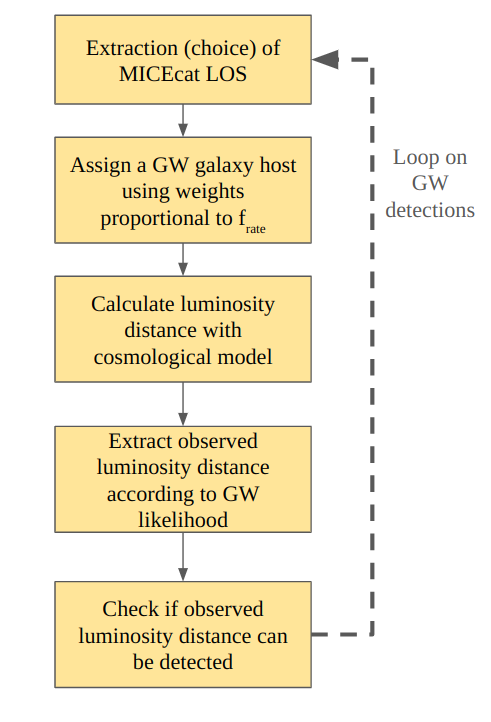}
    \caption{Flowchart of the process to simulate GW events from \MICE.}
    \label{fig:flowchart}
\end{figure}
In Fig.~\ref{fig:flowchart} we summarize the steps taken to obtain the GW catalogue from \texttt{MICEcat}. 

As argued in Sec.~\ref{Sec::Bayesian_Framework}, we perform our study by considering several sky localization areas for the GW events, hence the first step of the simulation is to filter all the galaxies reported by \texttt{MICEcat} in a given sky location. To do so, we use the python package \texttt{healpy} \cite{Zonca2019} to divide the sky in equal-sized pixels with the same sky area and then select all the galaxies falling in each of these line-of-sights (LOSs) considered.
For this study, we consider sky localization areas of 1 {\degs}, 10 {\degs} and 100 {\degs} to mimic the typical sky localizations of a network made by 3 GW detectors \cite{2018LRR....21....3A}. When dealing with GW events localized within 1 {\degs} and 10 {\degs}, we consider 100 pixels randomly distributed in the sky octant, while in the case that we are working with pixels of 100 {\degs} we subdivide the sky into 60 pixels.

After a LOS is selected, we apply a subsampling of all the galaxies reported by \texttt{MICEcat}. The resampling procedure is applied to speedup the computation of the hierarchical likelihood in Eq.~\ref{eq:final}, this is required due to the extensive number of runs that we have performed. We have verified that about $\sim 3000$ galaxies for each line of sight are enough to preserve the large-scale structures in redshift from which the Hubble constant is estimated, see App.~\ref{app:a} for more details. After the galaxies have been subsampled, we calculate for each of them a relative weight using Eq.~\ref{eq:pcbc} to host a GW events.

Using the calculated weights, we extract a galaxy as an host of the GW event and we compute its luminosity distance $d_L$ using the true cosmological model. Then we extract a value of the ``observed'' luminosity distance $\hat{d}_L$ using the GW likelihood model in Eq.~\ref{Eq::SOLikelihood}. If the observed luminosity distance is lower than a detection threshold $\hat{d}^{\rm thr}_L$, then we consider the signal as detected. 

We iterate the procedure described in this section until the desired number of GW sources is obtained.

\subsection{Properties of the GW catalogs}
\label{SubSec::Sub_Cataloghs}

\begin{figure}
    \centering
    \includegraphics[width = 0.5\textwidth]{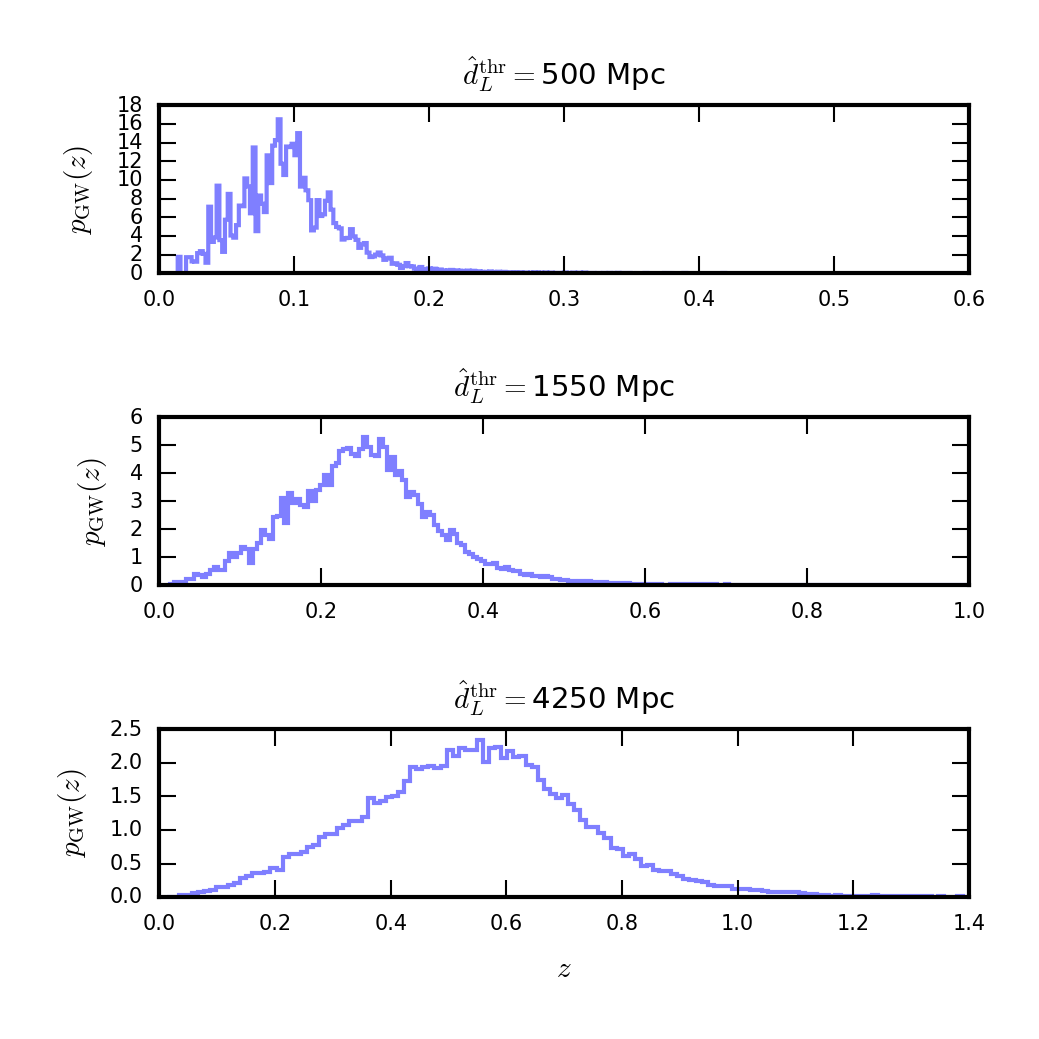}
    \caption{The histograms report the distribution of detected GW events, for different thresholds on the observed luminosity distance and number of detected events. The sky area considered corresponds to a sky-localization of 1 ${\rm deg}^2$ as an example.}
    \label{Fig::Detected_GW_dist}
\end{figure}

% \begin{figure*}[t!]
%     \centering
%     \includegraphics{images/Detected_Events_Smoothed_Histograms.png}
%     \caption{The histograms report the distribution of detected GW events, for different thresholds on the observed luminosity distance and number of detected events. The sky area considered corresponds to a sky-localization of 1 ${\rm deg}^2$ as an example. \SMc{Simplify as written in my comment in the text, also add the luminosity distance for the true cosmology on the top axis of the plot. Put Mpc on \dthr, use the cumulatives.}}
%     \label{Fig::Detected_GW_dist}
% \end{figure*}

Fig.~\ref{Fig::Detected_GW_dist} shows the distributions of the detected GW events for the three detection thresholds. The first property of the detected population is that the true redshift (luminosity distance) of the sources can extend beyond $\hat{d}_L^{\rm thr}$. This is because the GW likelihood model can introduce noise fluctuations that make the signals detectable. From Fig.~\ref{Fig::Detected_GW_dist}, we can also observe that a $\dthr=500$ Mpc would result in a population of GW events typically detected within redshift 0.2, $\dthr=1550$ Mpc within redshift 0.4 and $\dthr=4250$ Mpc within redshift 1. None of our simulations contains GW sources from galaxies close $z \approx 1.4$, which is the cut used to generate {\MICE}. As such, our studies do not need to account for the presence of this hard cut in the galaxy distributions.

\begin{figure}
    \centering
    \includegraphics{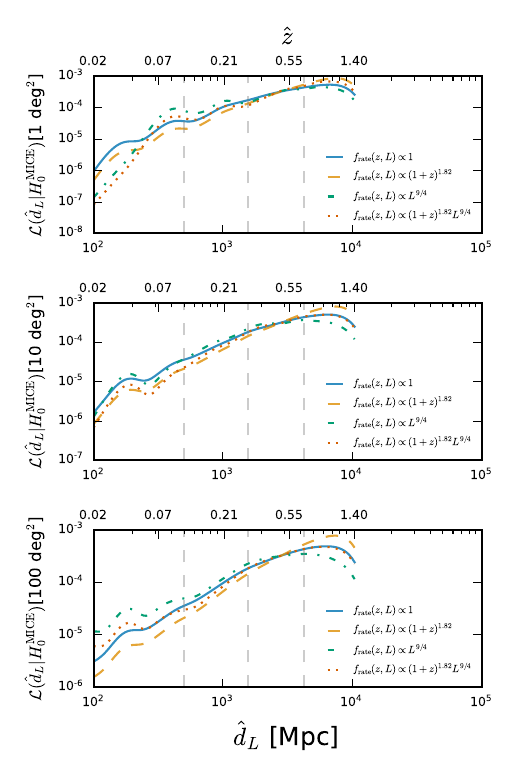}
    \caption{Single-likelihood estimator (c.f. Eq.~\ref{eq:final_single}) on the vertical axes as a function of the observed luminosity distance on the horizontal axis. The top, middle and bottom panels are generated assuming that the GW event is localized in a 1 {\degs}, 10 {\degs} and 100 {\degs} sky area. The different lines are generated assuming the four prescriptions of {\frate} used in this work. The figure has been generated assuming a \dthr=500 Mpc (we note that such a choice just acts as an overall normalization costant).} 
    \label{fig:likelihood}
\end{figure}

To quantify better how GWs detected at various distances interact with the construction of the galaxies' overdensity in the line-of-sight, we compute the single event likelihood
\begin{equation}
        \mathcal{L}(\hat{d}_L|H_{0}^{\rm MICE}) \equiv \frac{\sum_i^{N_{\rm gal}} \mathcal{L}_{\rm GW}(\hat{d}_L|d_{\rm L}(\hat{z}_g^i,H_0))f_{\rm rate}(\hat{z}_g^i,M(\hat{m}^i_g,H_0,\hat{z}_g^i))}{\sum_i^{N_{\rm gal}} P_{\rm det}^{\rm GW}(\hat{z}_g^i, H_0)f_{\rm rate}(\hat{z}_g^i,M(\hat{m}^i_g,H_0,\hat{z}_g^i))}\,.
        \label{eq:final_single}
\end{equation}
as a function of the observed distance $\hat{d}_L$ and conditioned on the true value of $H_{0}^{\rm MICE}$ used to generate {\MICE}. The single-likelihood profile can be used to understand how sensitive is the analyses to changes in the prescription of \frate. If the likelihood profile as a function of $\hat{d}_L$ is not modified as the prescription of $\frate$ changes, then the $H_0$ posterior is not expected to differ. We note that what impacts on the estimation of $H_0$ are changes in the trend of the likelihood and not its overall normalization factor.
In Fig.~\ref{fig:likelihood} we show this estimator calculated for three different LOSs areas. As we can see from the plots, GW events observed at low distances for small sky areas display large likelihood changes when the galaxy's luminosity is included in {\frate}. This is because when few galaxies are included in the localization volume, a change in the GW host probability would systematically prefer few galaxies, while if many galaxies are included there is no such preference. Instead, a change of prescription of {\frate} as a function of redshift does not affect the single-likelihood as this weight at low redshift is equal for all the galaxies. At higher distances (redshift), the redshift weight in {\frate} causes a more negligible change in the trend of the single event likelihood.

\section{Forecasting systematics on the $H_0$ inference}
\label{Sec::Results}

In this section, we present an extensive study on the systematics that could be introduced in the estimation of $H_0$ when mismodeling the GW host probability, namely the rate function $f_{\rm rate}$. To do so, we simulate 100, 200 and 500 GW detections from {\MICE} using the fiducial model for $f_{\rm rate}$ in Eq.~\ref{eq:fMR} and we then perform the inference using 4 different models for $f_{\rm rate}$. The first is the correct one and accounts for both redshift and absolute magnitude dependence, the second neglects the absolute magnitude, the third the redshift dependence and the fourth both the absolute magnitude and redshift. In estimating the $H_0$ we use a uniform prior in the range $[40,120]$ \hu.

For each of these cases, we build probability-probability (PP) plots (e.g., ~\cite{10.1093/biomet/55.1.1}) for the $H_0$ posterior. PP plots are built by repeating each simulation 100 times, each time drawing a value for $H_0$ in the prior range $[40,120]$ \hu. For each posterior obtained, we compute at which credible interval $\mathcal{I}$ the true value of $H^{\rm inj}_0$ is found, namely
\begin{equation}
    \mathcal{I}_i = \int_{40 \hu}^{H^{\rm inj,i}_0} p(H_0|\{x\}) dH_0.
\end{equation}
If no bias is present, the distribution of the $\mathcal{I}_i$ is expected to be uniform in the range [0,1] and its cumulative (the PP-plot) is expected to be a bisector in the range [0,1].

\subsection{Case 1: All GW sources are observed in the same line-of-sight}

\begin{figure*}[t!]
    \centering
    \includegraphics[width=
\textwidth]{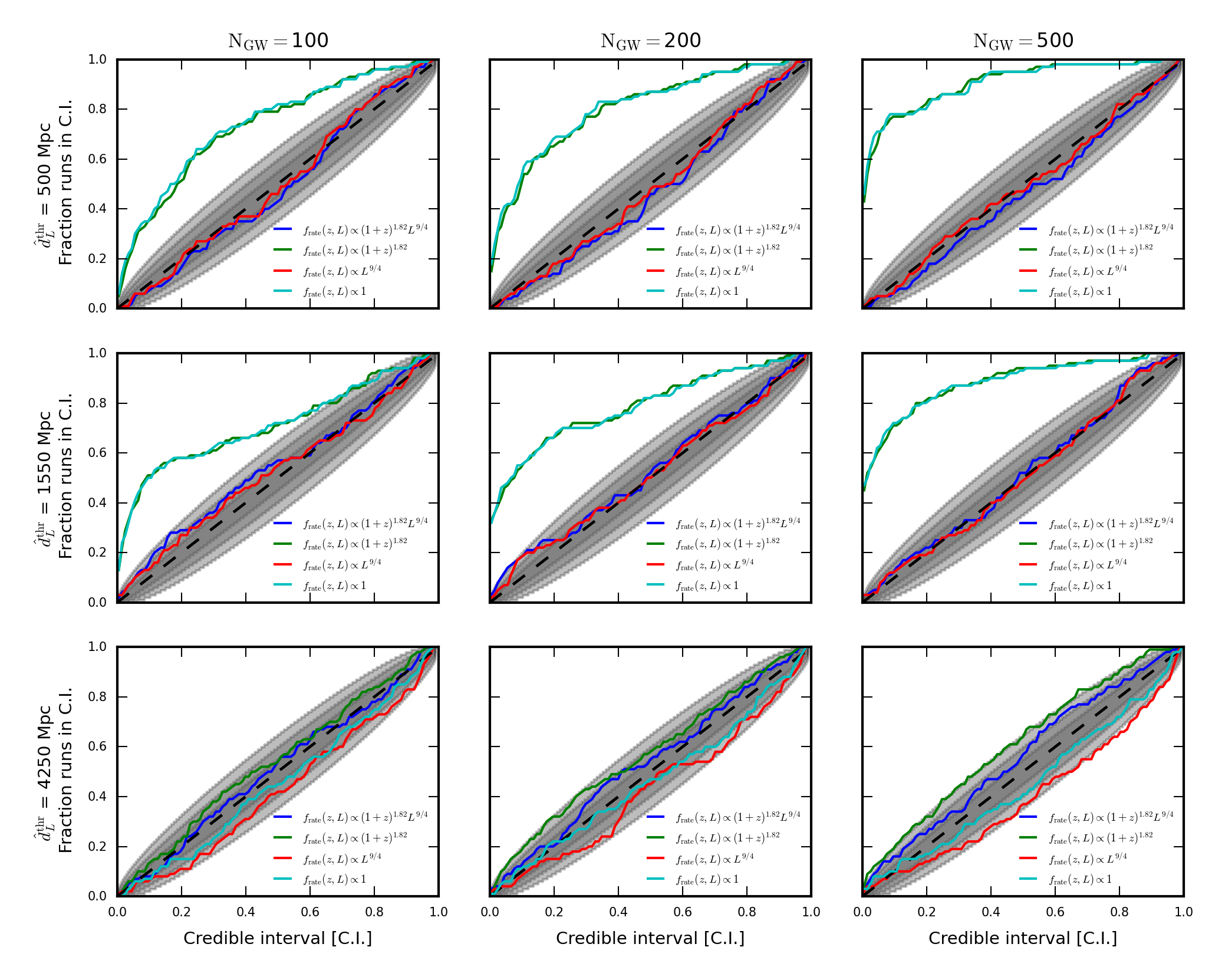}
    \caption{PP Plots corresponding to a single line of sight covering a sky-area of 10 $deg^2$. The rows correspond to the three different thresholds in luminosity distance, the columns to different number of detected GW events $N_{\rm GW}$.}
    \label{Fig::PP_Plots_Single_Line}
\end{figure*}

\begin{figure}[t!]
    \centering
    \includegraphics[width= 0.5
\textwidth]{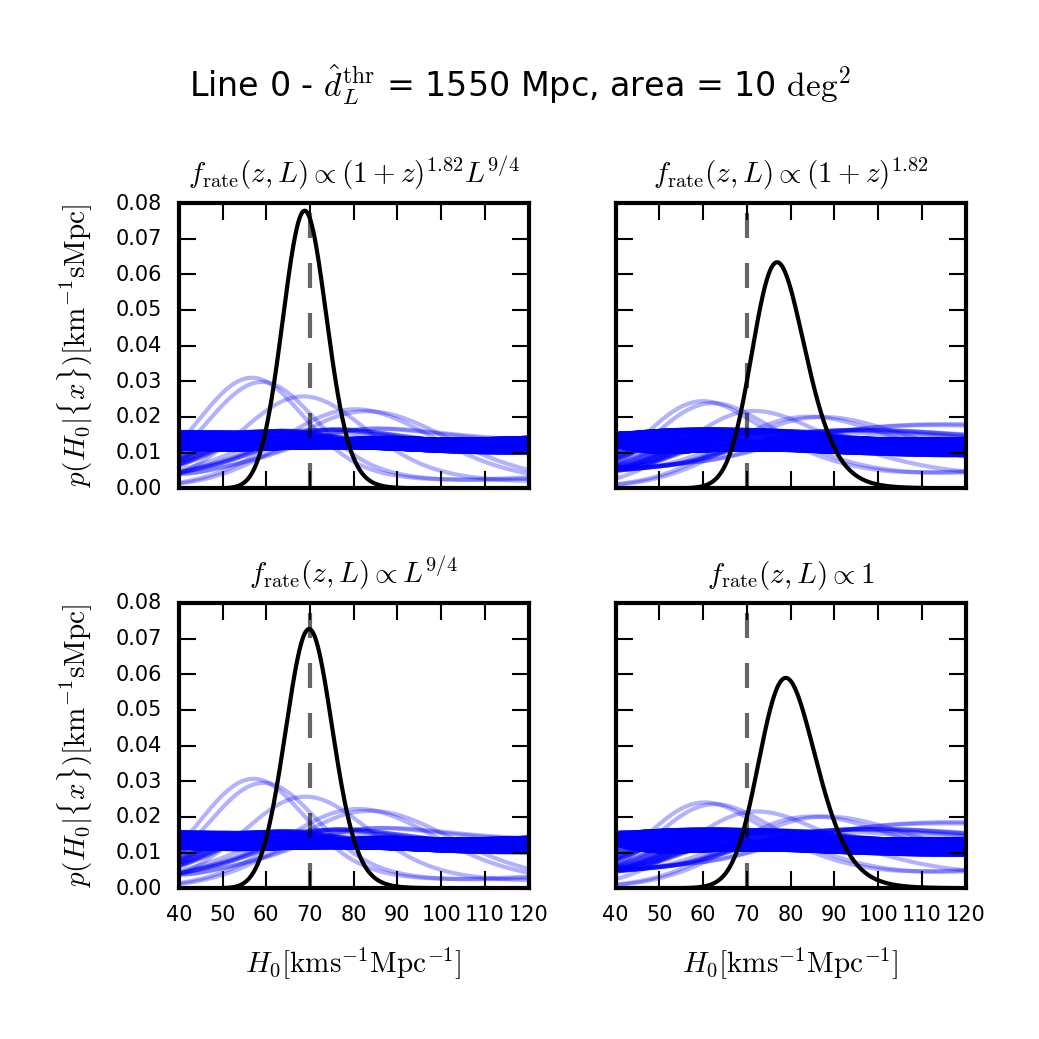}
    \caption{Posterior obtained fixing $H_0$ to 70 km $\rm{s}^{-1}$$\rm{Mpc}^{-1}$, $\hat{d}_L^{\rm{thr}} = 1550$ Mpc and the 200 detected GWs, for the same line of sight and a sky area of 10 $\rm{deg}^2$.}
    \label{Fig::Posterior_70_Single}
\end{figure}

We first start by discussing a simple case in which all the GW detections are obtained from the same LOS. In this limiting case, the distribution of galaxies, from the GW point of view, will strongly deviate from a uniform in comoving volume distribution. This is due to the fact we are observing GWs from the same LOS that contains always the same anisotropies in the redshift and luminosity distribution of galaxies. Thus we expect that systematically mismatching {\frate} for the same set of galaxies might introduce a systematic bias on the $H_0$ estimation.

In Fig.~\ref{Fig::PP_Plots_Single_Line} we show the PP plots for the various cases that we simulated using a sky localization area of 10 \degs. In all the cases, we obtain that when {\frate} is modelled correctly, the inference does not show any significant bias, as expected. 
However, if we consider a detection horizon $\leq 1550$ Mpc, i.e. we are observing GW events at small distances, we obtain that mismodelling  {\frate} can introduce a significant bias. In particular, we found that if galaxy luminosities are neglected, then the $H_0$ posterior will typically display a bias towards \textit{higher} values of $H_0$ and this effect depends on the galaxy distribution along the LOS considered. In fact, from the plots in Fig.~\ref{Fig::PP_Plots_Single_Line}, we can also observe that this is the dominant source of the bias, even if the redshift dependence of {\frate} is neglected. The magnitude of the $H_0$ bias increases as more GW events are used for the inference as expected for a systematic bias stacking on each of the GW events analyzed. If we consider \dthr$= 4250$ Mpc, we find that mismodelling {\frate} does not introduce any significant bias. We verified that the results obtained for 10 {\degs} are also valid for 1 {\degs}, while for 100 {\degs} the bias results weakened since, when considering large volumes, the local anisotropy tends to be averaged out.

The results presented in the previous paragraph can be explained as follows. On one hand, when \dthr$<1550$ Mpc, we are typically dealing with close-by GW events whose luminosity distance uncertainties are ``small'' \footnote{remember that in our model the typical luminosity distance uncertainties are of the order of 20\% the luminosity distance}. As a consequence, they include in their localization volume galaxies distributed in a \textit{narrow} redshift space but can show some significant anisotropies in terms of luminosity distribution. Therefore, mismatching the dependence of {\frate} from the galaxies' luminosity introduces a strong bias for the inference while the effect of mismatching the redshift dependence is small. On the other hand, if we have a detection threshold \dthr$=4250$ Mpc, most of the GW events will be located at high distances and will have a large localization volume. As a consequence, they will include galaxies over a \textit{wide} redshift range but as the redshift range is large, we do not expect to see any significant anisotropy in their distribution in terms of luminosity. It follows that in this case, mismodelling the redshift dependence on {\frate} is more important than mismodelling luminosity dependence. Fig.~\ref{fig:likelihood} corroborates the above discussion, which shows that the single-event likelihood is different at low redshift when mismatching the luminosity weight and at high redshift when mismatching the redshift dependence. 

The behavior of $H_0$ is directly linked to the type of mismodelling that we are doing on {\frate}. Remembering that $H_0 \propto  z/d_L$, when a bias is introduced by a wrong modelling of the redshift dependence, we systematically prefer to set GW hosts at lower redshift compared to their true distribution. This will result in a bias towards lower $H_0$ values (hinted also by the curves in Fig.~\ref{Fig::PP_Plots_Single_Line}). When we neglect the intrinsic luminosity dependence of \frate, we are more likely to assign GW events to faint galaxies and for this particular LOS translates to a high $H_0$ bias. 
To provide a more qualitative picture of the biases, we show in Fig.~\ref{Fig::Posterior_70_Single} the reconstructed posterior for $d_L^{\rm thr}$ = 1550 Mpc and 200 detected GWs and with a sky area of 10 $\rm{deg}^2$, where the true value of $H_0$ has been set to 70 \hu. As it can been seen from the plot, the $H_0$ posterior is sensitive to the models of \frate. Although for this particular case, the statistical errors are sufficiently large to include the true value of $H_0$, the overall statistical framework suffers from systematic biases as demonstrated by the PP plots.

\subsection{Case 2: GWs from all the sky directions}

We now repeat the same studies of the section above by randomizing the LOS from which each GW event is observed. This procedure is equivalent to assuming that the GW distribution is isotropic in our Universe. In this case, we would expect the average distribution of galaxies in redshifts over all the LOSs to be reasonably uniform in comoving volume and thus, any local anisotropy present in the luminosity distribution should be averaged out. In other words, we would expect that when not accounting for the galaxies' luminosity in {\frate}, the systematic introduced on $H_0$ will be sometimes to the right and sometimes to the left, thus resulting in no bias. When we consider GW events detected at distances of $\mathcal{O}$(Gpc), mismatching the redshift dependence leads to a further systematic bias, shifting the posterior towards lower values of $H_0$. As explained in the previous section, when the redshift range considered is large, mismodelling the redshift dependence translates in preferring nearer galaxies as GW hosts, hence providing a shift of the posterior to the left.

\begin{figure*}[t!]
    \centering
    \includegraphics[width=
\textwidth]{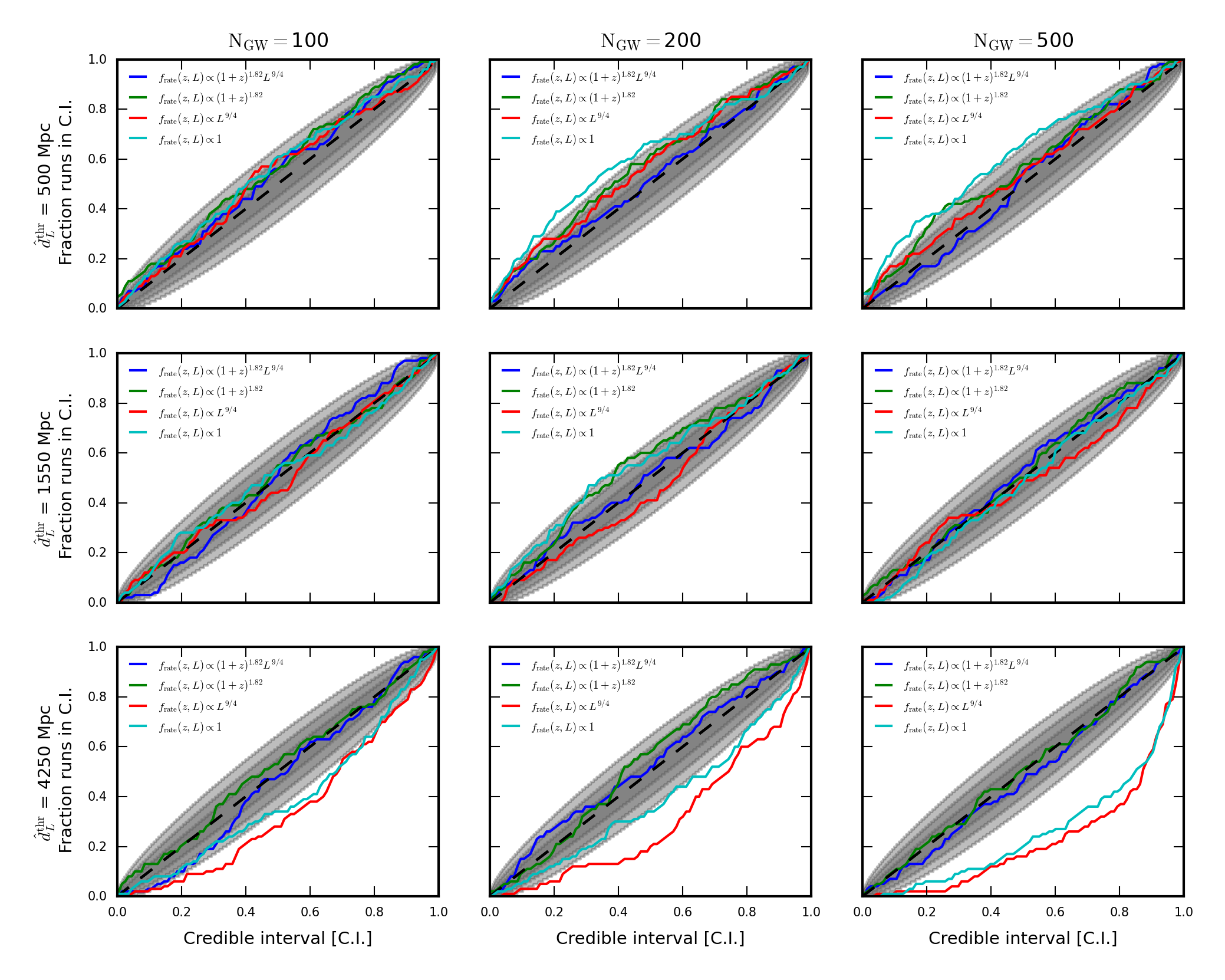}
    \caption{The figure reports the PP Plots obtained taking each time a different line of sight when estimating the posterior, each from the 100 ones chosen covering a sky area of 10 $deg^2$. The rows correspond to the three different thresholds in luminosity distance, the columns to different number of detected GW events $N_{\rm GW}$.}
    \label{Fig::PP_Plots_Multiple_Lines}
\end{figure*}

\begin{figure}[t!]
    \centering
    \includegraphics[width= 0.5
\textwidth]{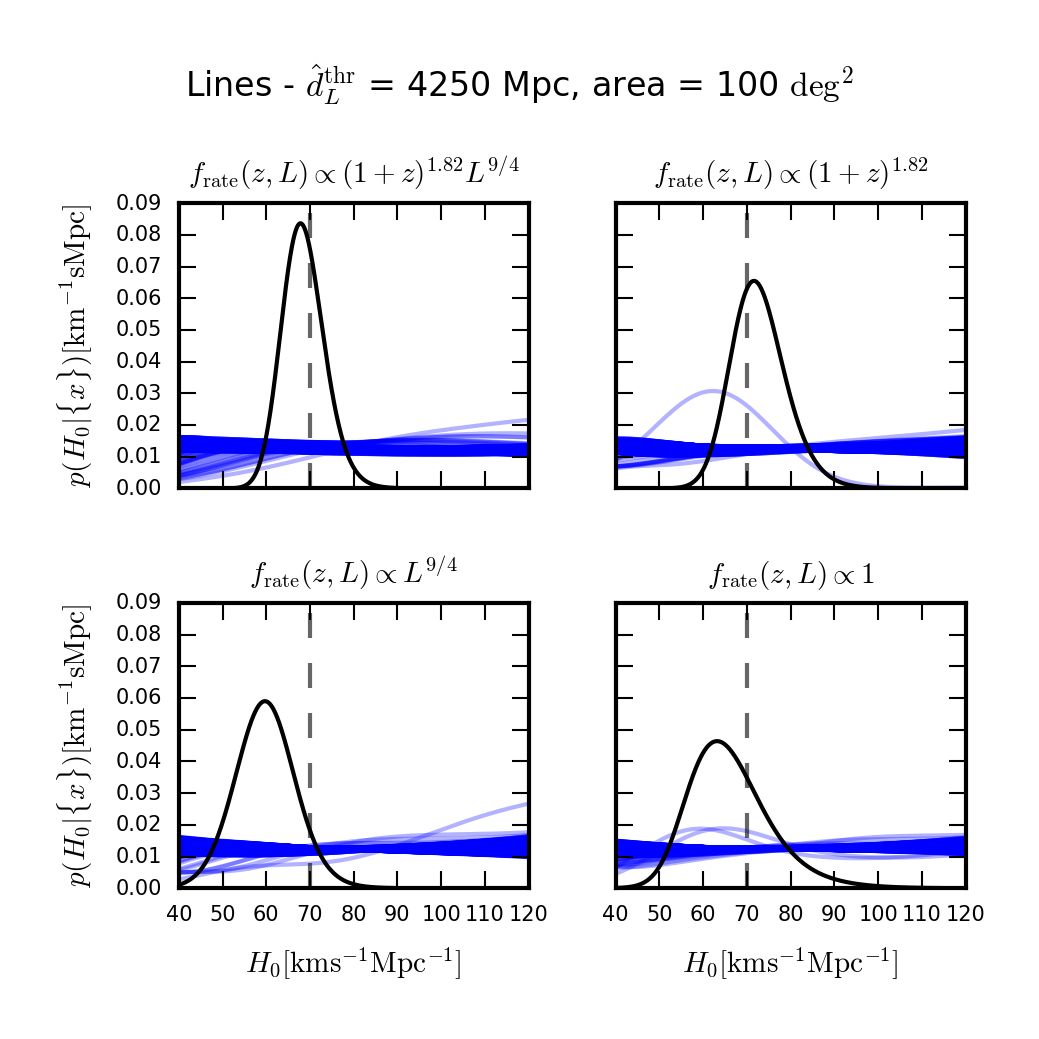}
    \caption{Posterior obtained fixing $H_0$ to 70 km $\rm{s}^{-1}$$\rm{Mpc}^{-1}$, $\hat{d}_L^{\rm{thr}} = 4250$ Mpc and the 500 detected GWs, for a sky area of 10 $\rm{deg}^2$ and considering different lines of sight when reconstructing the posterior.}
    \label{Fig::Posterior_70_Multiple}
\end{figure}

Fig.~\ref{Fig::PP_Plots_Multiple_Lines} shows the PP plots for a sky area of 10 \degs. Also in this case, we observe similar behaviours to the ones observed in the previous section: if the CBC rate function as a function of redshift if mismatched, this will result in a bias when considering GW events detected at higher distances. Differently to the single-LOS case, when the intrinsic galaxy luminosity is mismatched, we obtain the posterior on $H_0$ that is typically unbiased, even for close-by and well-localized GW events. This is consistent with the fact that by marginalizing over different LOSs, galaxies' clustering properties tend to follow a uniform comoving volume distribution that is unaffected by the actual galactic properties. We provide a qualitative picture in Fig \ref{Fig::Posterior_70_Multiple}, where we show the reconstructed posterior for $d_L^{\rm thr}$ = 4250 Mpc and 500 detected GWs and with a sky area of 10 $\rm{deg}^2$, where the true value of $H_0$ has been set to 70 \hu.  

\section{Conclusions}
\label{Sec::Conclusions}

In this work, we have performed a further systematic study on the bias that can be introduced on the inference of $H_0$ for dark sirens cosmology with galaxy catalogs. We performed extensive end-to-end simulations to infer $H_0$ from GW standard sirens starting from \texttt{MICEcat}, a simulated universe reproducing the large-scale structures properties of our Universe.

Assuming the galaxy catalogue to be complete and with a set of simplified assumptions for the GW detection probability and GW localization, we have shown that mismodelling of the CBC host probability, namely \frate, can introduce relevant biases when estimating $H_0$. We have found that mismatching the {\frate} dependence as a function of redshift will most likely introduce a systematic shift at low $H_0$ when dealing with hundreds of GW detections detected at luminosity distances higher than 2Gpc. We have also found that mismatching {\frate} as a function of the galaxy's luminosities might introduce a systematic bias on $H_0$ when dealing with multiple events from the same LOS detected at very close distances. However, we have shown for the first time that when dealing with GW events from different LOSs, even close ones, mismatching the galaxy luminosity dependence in {\frate} is not expected to bring significant bias on $H_0$.

The results presented in this paper are also scalable to other GW detector networks. If we assume that the luminosity distance errors do not strongly change with the SNR \footnote{This is a reasonable assumption if we assume that the luminosity distance inclination degeneracy dominates the GW error budget \citep{2019PhRvD.100h3514C}.}, then the luminosity distance detection thresholds could be rephrased in terms of SNR detection thresholds. Thus working with luminosity distance threshold would implicitly mean working with standard sirens selecting according to an SNR cut that depends on the detector network. As an example, a binary black hole binary at 500 Mpc with current GW detectors would enter the analysis with a SNR cut of $\gtrsim 40$, while for future generation detectors would enter the analysis for an SNR cut $\gtrsim 10^3$.

We warrant that these results are valid with some caveats. First, we assumed that more than 100 GW events were used for the $H_0$ inference. In the case of a low number of GW detections (i.e. few localized standard sirens with very high SNR), we expect that assumptions on the host galaxy properties might introduce a systematic bias of $H_0$ as on average the LSSs seen by the GW events are not uniform in comoving volume. In other words, a low number of highly localized GW sources would mimic the behaviour observed for the case of 100 well-localized sources drawn from the same LOS. Second, our results were generated assuming a complete galaxy catalog in terms of CBC hosts. If the galaxy catalog is not complete, a completeness correction (that strongly depends on \frate) \citep{mastrogiovanni2023novel} should be applied. It is currently unknown how mismatching the completeness correction would result in a systematic bias. We leave this analysis for a future project.

\begin{acknowledgements}
CloudVeneto is acknowledged for the use of computing and storage facilities. A.R. thanks Steven N. Shore for valuable comments on the draft and M. Cignoni for discussions. AR acknowledges financial support from the Italian Ministry of University and Research (MUR) for the PRIN grant METE under contract no. 2020KB33TP and from the Supporting TAlent in ReSearch@University of Padova (STARS@UNIPD) for the project ``Constraining Cosmology and Astrophysics with Gravitational Waves, Cosmic Microwave Background and Large-Scale Structure cross-correlations''.
\end{acknowledgements}

\appendix

\section{Subsampling of the line-of-sights}

\label{app:a}

\begin{figure}[t!]
    \centering
    \includegraphics[width=0.5\textwidth]{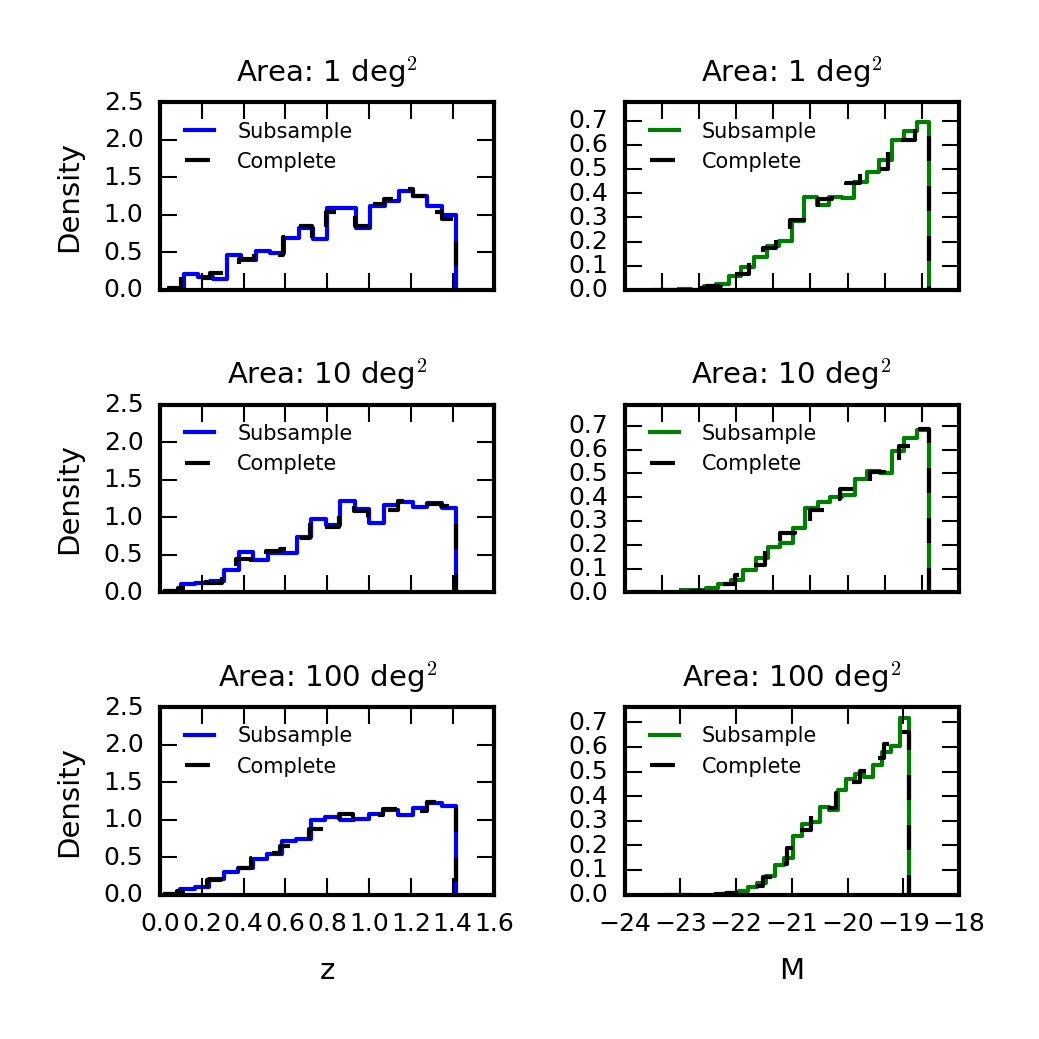}
    \caption{Comparison of the distribution of redshifts (left column) and magnitudes (right column) between original galaxies for some selected LOSs in {\MICE} and the subsampled ones. The rows report sky areas respectively of 1 \degs, 10 \degs and 100 \degs.}
    \label{Fig::Example_Histogram}
\end{figure}

To ease the computational burden of the analyses, we subsample the galaxies reported in every line-of-sight of MICEcat. We have found empirically that a subsampling of about $\sim3000$ galaxies does not modify the statistical properties of the galaxies' distribution. 
We report in Fig.~\ref{Fig::Example_Histogram} the comparison between the redshift (blue) and absolute magnitude (in orange) for the subsampled data with the full data (black) for the different sky areas we consider. We quantify the difference between the original and subsampled distributions using the discrete Kullback-Leibler (KL) divergence calculated by subdividing the redshift and magnitude ranges in 20 bins. 
We have found a KL divergence to be at most $~10^{-1}$. Hence we considered the sumbsampled and the full distributions to be statistically indistinguishable.

\bibliographystyle{aa} % style aa.bst
\bibliography{Bibliography}
\end{document}